\documentstyle[preprint,tighten,eqsecnum,epsfig,aps]{revtex}
\input{psfig}

\def\beq{\begin{equation}}   \def\eeq{
\end{equation}}

\begin{document}
%\draft

{\normalsize {\large }}

%%%%%%%%%%%%%%
\title{
On the increase with relative distances of light cone operator
product of currents and related phenomena}
\author{ B. Blok$^1$\thanks{E-mail:blok@physics.technion.ac.il}
 and L. Frankfurt$^2$\thanks{E-mail:frankfur@lev.tau.ac.il} }
\address{$^1$Department of Physics, Technion -- Israel Institute of
Technology, Haifa 32000, Israel\\[10pt]
 $^2$School of Physics and
Astronomy, Faculty of Natural Sciences, Tel Aviv University,
Israel.} \maketitle

\thispagestyle{empty}

\begin{abstract}
We show that in QCD in the leading twist approximation flavor singlet light cone 
current-current correlators increase with distances rapidly and without spatial 
oscillations. 
 \end{abstract}

 \setcounter{page}{1} \section{Introduction}

\par It has been suggested that increase with energy of cross sections of hard processes 
found in QCD is equivalent in coordinate space to the increase with distance 
of flavor independent correlators of currents on light cone ref. \cite{BF} . A 
number of simplifying approximations made in ref. \cite{BF} like
double logarithmic approximation with fixed coupling constant precludes
unambigous conclusions.  In this paper we show that increase of
correlators of currents with distance is valid in the DGLAP,  BFKL and
resummation  approximations. Space-time oscilations of correlators of
currents which are present in the parton model \cite{Ioffe}, disappear
in QCD for sufficiently large invariant length $py$.

\par Increase with energy of cross sections of hard processes can not continue
forever because it contradicts to conservation of probability.
\cite{AFS} Application of LO DGLAP approximation to the
interactions of colorless gluon dipole $gg$ with the  virtuality
$Q^2\approx 10 GeV^2 \gg Q_o^2$ with a nucleon target reveals this
conflict with probability conservation in the kinematics of
$x=(Q^2/W^2+Q^2)\le x_{cr}(Q^2) \approx 10^{-5}$ achievable at
LHC.\cite{AFS,ANRS}.($Q^2\approx 10 GeV^2$ was chosen to guarantee
smallness of running coupling constant.) At the same time the
amplitude of the scattering of $q\bar q$ dipole in this kinematics is far from S matrix 
unitarity limit.  Onset of new pQCD regime at $Q^2 \approx 1-2 GeV^2$ for
the interaction of $q\bar q$ dipole has been proposed in \cite
{A.Mueller,McLerran} and references therein.

\par Account of unitarity of S matrix, of energy-momentum conservation
law and nonconservation of bare particles helps to show that limiting behavior of 
hard processes at sufficiently high energies is described by black disc with the 
radius increasing with energy=BDL regime.\cite{McDermott,ANRS}. A lot of new hard
phenomena which may appear effective tool to identify BDL regime can be observed at 
LHC . For the review and references see \cite{ANRS}.  The aim of this paper is to 
demonstrate that color and flavor blind correlators of currents calculated within
conventional approximations rapidly increase with relative distance. Such a 
behavior of correlators is due to Lorentz dilatation of life time of colorless dipole 
produced by current in the target rest frame. It resembles turbulence where correlators
of velocities are increasing with relative distances \cite{LL}. Increase with distance of the 
correlators calculated in LT approximation and even faster increase with distances of HT 
effects suggests similarity between  the transition to BDL and phase transition.  Really it is 
well known that key property of a system near critical point for a second
order phase transition is increase of correlators with
distance\cite{LLSM}. Since our interest is in the dependence of
commutator of currents on distance but not its absolute value we
will not distinguish below between different correlators
(different structure functions).

\par For definitness we begin our discussion from the DIS lepton scattering
off hadron target T: $e+T\to e+X$. In this case $Q^2$ is the
virtuality of photon and $W$ is invariant energy of $\gamma^* T$
collision. Our interest is in the regime of small $x=Q^2/W^2+Q^2$
which is however significantly larger than $x_{\rm cr}(Q^2)$
characteristic for BDL. It is well known that total cross section
of DIS can be calculated as matrix element of correlator of
electromagnetic currents at light-cone . Thus knowledge of
structure functions as a function of $x,Q^2$ is sufficient to
reconstruct correlators of e.m. currents in coordinate space.
B.Joffe \cite{Ioffe} was the first to extract this correlator from
experimental data on structure functions of DIS within the
framework of parton model. In ref. \cite{Frishman} Fourier
transform into coordinate space of variety of expressions for
structure functions has been calculated.In this paper we restrict
ourselves by evaluation of commutator of currents at large
invariant distances $(py)\to \infty$ and small but fixed $y^2$.
(Here $y$ is relative space-time difference between currents
within a correlator.) In this kinematics correlator is dominated
by structure function in the kinematics $x\ll 1$ and small but
fixed $Q^2$.

\par The method of moments or Wilson operator expansion for the product
of currents is convenient to evaluate the behavior of the
current-current commutator in the coordinate space.
\begin{eqnarray}
<N\vert j_{\mu}(y)j_{\nu}(o)\vert N> &=& (1/y^2)^2\sum_n
p_{\mu}p_{\nu} (py)^n <N\vert O_n(0)\vert N> +\mbox{NLT
terms}\nonumber\\[10pt]
 &=&p_{\mu}p_{\nu} (py)f(py,y^2)/(y^2)^2 +\mbox{NLT
terms}\nonumber\\[10pt]
\label{fr1}
\end{eqnarray}
where (py) is a kinematical factor.
In this paper we shall work with 
\beq
G(py,y^2)=\int d^4q G(x,q^2)\exp(iqy)
\label{de1}
\eeq
where $G(x,Q^2)$ is a structure function.

\par Another useful quantity is the Fourier transform of gluon density
which can be interpreted as $\propto $ effective coloress  "potential" 
$D$ of dipole-target interaction. 
\beq
D(py,y^2)=\int d^4q \frac{G(x,q^2)}{2pq}\exp(iqy) 
\label{11} 
\eeq
This "potential" is a Fourier transform of a dipole-target cross-section, and 
$xG$ is a corresponding gluon structure function. (We use here conventional 
notations for structure functions. The name "potential" is used because within 
the nonrelativistic approximation $D$ has meaning of potential.)  $D$ is often 
used in the eikonal models for the structure functions in the regime of small $x$ 
as "potential" \cite{A.Mueller,McLerran})

\par We have also  studied the closely related problem of coherence length in
QCD. We explain that coherence length which measures longitudinal length of 
virtual photon wave function in the target rest  frame is  $l_c \approx 1/2m_Nx$ , 
far from BDL. However near BDL  coherence length
 becomes $\propto x^{1-\mu}\ll 1/2m_{N}x $ because
 of a more rapid increase with energy of the interaction of smaller dipole with a target.
$\mu \approx 0.2$ as known from theoretical and experimental investigation of 
$xG_{p}(x,Q^2)$.   This is because of the  enhancement of contribution of small 
transverse size configurations in the wave function of photon whose interaction 
increases with energy faster than for average configurations.

\par Our results suggest that near BDL the dipole wave function should be
modified due to the long range dipole-target interactions discussed
above. Significant change of the wave function occurs in the kinematics
where  leading twist approximation is violated.

\par The paper is organized in the following way. In section 2  we consider
the correlators of currents in the leading log DGLAP approximation. Then we 
explain that similar behaviour arises in other pQCD approaches also. In section 3 we
discuss the coherence length in QCD far and near BDL. In section 4 we discuss 
the deformation of the dipole wave function. Our conclusions are in the section 5.

\section{Current-current correlators in the leading log DGLAP approximation.}

\par The DGLAP evolution equation in the leading log approximation is
\beq Q^2\frac{d}{dQ^2}G(x,Q^2)=(\alpha_s/(2\pi ))\int^1_x(dx'/x')
\gamma_{G}(x/x')G(x',Q^2). \label{1} \eeq Here $\gamma_{G}$ is
the kernel in the QCD evolution equation.
\par Consider first the case of a freezed coupling constant
(considered in ref. \cite{BF}). In this case the structure function is
given by 
 \beq G(x,Q^2)= \int_C
\frac{dj}{(2\pi i)} x^{-j}D(j,\alpha_s\log(Q^2/Q^2_0))\label{90a}\eeq
The contour of integration over j (we denote it C) runs along a straight
line parallel to the imaginary axis to the right of all singularities of
the integral. We use the notation $Q^2=-q^2$ if $q^2\le 0$ and $Q^2=q^2$
if
$q^2\ge 0$. The Bjorken scaling variable is defined in a usual way:
 \beq
x=-q^2/(2pq).
\label{3}
\eeq
\par For the
anomalous dimension we have: \beq
V_{\pm}(j)=0.5(V_F(j)+V_G(j)\pm\sqrt{(V_F(j)-
V_G(j))^2+24\Phi^G_F(j)\Phi^F_G(j)},
\eeq where \beq V_F(j)=-C_2(4\psi(j+1)+4\gamma_E-3-2/(j(j+1)),
\eeq \beq
V_G(j)=-4N_c(\psi(j+1)+\gamma_E)+11N_c/3-2+8N_c(j^2+j+1)/(j(j^2-1)(j+2)),
\eeq \beq \Phi^G_F(j)=2C_2(j^2+j+2)/(j(j^2-1)), \eeq \beq
\Phi^F_G(j)=(j^2+j+2)/(j(j+1)(j+2)). \eeq The function $\psi(j)$
is the usual logarithmic derivative of Gamma function
$$\psi(j)=\frac{d\Gamma (j)}{dj}$$
and
\beq
\gamma_G(j)\equiv V_+(j)
\label{fg}
\eeq
\par Let us begin  calculations within the simplifying assumption of
frozen coupling constant
and then to analyse realistic case. It is easy
to carry out the integration over $d^4q$ in the Fourier transform.
We use the formulae \cite{BF}: \beq (q^2)^\beta F(x)\rightarrow
\pi\Gamma(\beta +2)\frac{(py)^{\beta +1}}{(y_+^2)^{\beta+2}}
\int^1_0F(x)x^\beta\cos(x(py)+\beta\pi/2)dx \label{fur} \eeq if
the latter integral converges. Here $u_+=\theta (u)u$. For the
case of the power like behavior of $F(x)$ this integration can be
performed analytically, and in the limit of large invariant length
$(py)\gg 1$ one obtains power like asymptotics: \beq (q^2)^\beta
x^\alpha\rightarrow \pi\Gamma (\beta
+2)\Gamma(\alpha+\beta+1)\cos((2\alpha+\beta)\pi/2)/((py)^\alpha)(y^2)^{\beta
+2}) \label{ft} \eeq In our case $\beta=\gamma(j),\alpha =-j$, so
\beq G(y^2,py)=(\pi/2)\int dj
(py)^{+j}\frac{1}{(y^2)^{\gamma(j)+2}}
\Gamma(\gamma_G(j)+2)\Gamma(1-j+\gamma_G(j))\sin((-j+2\gamma_G(j))\pi/2)
\eeq
\par We now use the saddle point method to integrate over j.
\beq
\frac{\log(y_0^2/y^2)}{\log(py)}=\alpha_s\frac{d\gamma_G (j)}{dj}
\label{skuns}
\eeq
The saddle point can be found  numerically. The derivative of the DGLAP
anomalous dimension can be approximated with good accuracy by
$$ \gamma'(n)\sim -12/(n-1)^2$$ for $1\le n\le 2.5$ and
$$\gamma'\sim -5/(n-1)$$ for $n\ge 3$. This means that the double log
expression works quite good numerically for
\beq
\log (py)\ge \alpha_s\log(y_o^2/y^2)
\label{v}
\eeq
Using the latter numerical approximations for anomalous dimension,
one obtains the relevant saddle point:
\beq
j=1+\sqrt{12\alpha_s\log(y^2_0/y^2)/\log(py)}
\label{new}
\eeq
i.e. we have the double log expression. However now we are permitted to
take
into account also the terms that contain only one big logarithm.
For $\log(py)\ge \alpha_s\log(1/y^2)$, one has
\begin{eqnarray}
G(py,y^2)&=&\frac{\log(Q_0^2y^2)^{1/4}}{\log(py)^{3/4}}\frac{1}{(y^2)^2}
\exp\big[\sqrt{(4\alpha_sN_c/\pi)\log(py)\log(Q_0^2y^2)}\big]
\nonumber\\[10pt]
&\times&\cos\{(\sqrt{(\alpha_sN_c/\pi)}(\sqrt{\log(Q_0^2y^2)/\log(py)}+
2\sqrt{\log(py)/\log(Q_0^2y^2)})\pi/2\}\nonumber\\[10pt]
&\times&\Gamma\{\sqrt{(\alpha_sN_c/\pi)}(\sqrt{\log(Q_0^2y^2)/\log(py)}+
\sqrt{\log(py)/\log(Q_0^2y^2)})\}\nonumber\\[10pt]
&\times&\Gamma\{2+
\sqrt{(\alpha_sN_c/\pi)(\log(py)/\log(Q^2_0y^2)}\}\nonumber\\[10pt]
\label{gru}
\end{eqnarray}
\par Consider now the realistic case of the running coupling constant.
In this case \cite{DGLAP,DDT,Doc} the gluon structure function is
given by
 \beq G(x,Q^2)= \int_C
\frac{dj}{(2\pi i)} x^{-j}D(j,\xi)\label{90}\eeq
 Here \beq
D(j,\xi)=\frac{(V_+(j)-V_0(j)}{V_+(j)-V_-(j)}\exp(V_+(j)\xi),
\label{91} \eeq where \beq
\xi=\frac{1}{b}\log(\alpha_s(Q_0^2)/\alpha_s(Q^2)), \eeq
and
\beq
b=11-\frac{2}{3}N_F=9\,\, ,\,\, N_F=3
\label{k1}
\eeq

The calculations one needs to carry through for  the realistic case of the
 running structure constant are completely similar to the 
calculations with the freezed coupling constant. The only change in
 the calculation procedure is the substitution in all 
calculations of the terms like $\log(Q^2/Q_0^2)$
 by $\log(y_0^2/y^2)$. Consequently, we 
have to substitute $\alpha_s\log(y_0^2/y^2)$ by
\beq
\xi(y^2)=\frac{1}{b}\log(\alpha_s(1/y^2)/\alpha_s(1/y_0^2)).
\label{sp}
\eeq
The saddle point will be in the point
\beq
j=1+\sqrt{\frac{12\xi(y^2)}{\log(py)}}
\label{fp}
\eeq
and we obtain
\begin{eqnarray}
G(py,y^2)&=&\frac{\log(Q_0^2y^2)^{1/4}}{\log(py)^{3/4}}\frac{1}{(y^2)^2}
\exp(\sqrt{12\xi(y^2)\log(py)})\nonumber\\[10pt]
&\times&\cos((\sqrt{\xi(y^2)/\log(py)})
\times\Gamma(((\sqrt{\xi(y^2)/\log(py)})
)\nonumber\\[10pt]
\label{grur}
\end{eqnarray}
Similar calculations can be carried for an opposite case
$\log(py)\le \alpha_s\log(y_o^2/y^2)$, and one can easily see that
the correlator increases with the increase of the invariant length
in this case also.
\par Similarly it is easy to derive the increase with distance of the effective
dipole-target potential.
\par One of the important features of the correlators in the parton model
is their space-time oscillations \cite{Ioffe}.We will show that
oscillations at large distances in the correlator and "potential" for
dipole -target interactions disappear in pQCD.
\par For fixed $y^2$ and frozen coupling constant the correlators and 
the effective dipole-target potentials derived above oscillate, as shows the 
existence of the $\cos$ term.
 The period of oscillations can be determined from the
condition that the argument of the cosinus is $n\pi$ where $n$ is
the integer number.
\beq 
\gamma(j_0)+j_0= n 
\eeq 
where $j_0$ is the saddle point. We immediately obtain the condition 
\beq
\sqrt{\alpha_sN_c/\pi}(\sqrt{\log(1/y^2)/\log(py)}
+\sqrt{\log(py)/\log(1/y^2)})=n 
\label{no} 
\eeq 
In our approximation the second term is bigger than the first and we
obtain that the oscillation period is 
\beq 
py\sim (1/y^2)^{n^2/(\alpha_sN_c/\pi)} 
\label{no1} 
\eeq 
Thus oscillations exist in the case of frozen coupling constant for large py. 
This analysis can be easily upgrated to
analyze the case of the running coupling constant. In this case
the relevant condition is 
\beq 
\xi(y^2)/\log(py)=n^2 
\label{po}
\eeq 
Hence 
\beq 
py\sim \exp(\xi(y^2)/n^2) 
\label{ui} 
\eeq 
Thus taking into account of running coupling constant changes
dramatically the oscillation period: instead of increasing with n
it now decreases. Therefore oscillations are present at small (py)
which are beyond of region of applicability of our consideration.

\par The reason for such dramatic influence of asymptotic freedom  is
evident in the procedure used to calculate the integral over
$q^2=-Q^2$. This integral tis the product of rapidly
oscillating $\cos(q^2(t-r)/x_B)$ and $F(q^2)$  \cite{Ioffe} 
\beq
G(px,x^2)=-\frac{\pi}{2m(py)}\int dq^2\int
dx_BG(x_B,Q^2)(q^2/x^2_B)\sin{(q^2/(m^2x_B)x^2/(py)+(py)x_B)}
\label{1371} 
\eeq 
In the case of frozen coupling constant  the function G behaves as the power of $Q^2$, 
and rapidly changes within the period of oscillations of cosinus. Thus exact
calculation of the integral produces terms proportional to $\cos(V(j)\pi/2$, or corresponding 
sinus because of  $F(Q^2)\sim Q^{2V(j)}$. This gives rise to the above result . However in 
the case of the running coupling constant the function $F(Q^2)$ depends on $Q^2$
more slowly , due to the presence of the additional logarithm. Indeed, the characteristic period of
oscillations is $\sim x(py)/y^2$. However the characteristic $x$ is connected to 
characteristic $(py)$ (coherence
length) as $py\sim 1/(x)^\beta$ where $\beta$ is a number close
but smaller than one. Then  the period oscillations is $\sim
(py)^{1-1/\beta}/y^2$. Thus the period of oscillations decreases
with increasing $(py)$. Evidently , the function $\xi(y^2)$ on
this scales can be considered constant, and one can safely carry
out $\xi(q^2)$ from the integral and substitute it by $\xi(y^2)$.
The oscillations die down for running coupling constant and large
$(py)$. The behavior for the case of $\log(py)\le \alpha_s\log(1/y^2)$
is qualitatively similar. Thus the  oscillations we found within
the DGLAP approach are qualitatively different from the
oscillations in the parton model found in ref.\cite{Ioffe}.
\par Similar analysis can be carried through for the BFKL case
\cite{BFKL} by substituting $q_o=1/(t-z)$ in the Fourier transform
as it is legitimate in the leading $\alpha_s\ln(x_0/x)$
approximation.
\par Formally, we use for the BFKL case the contour integral representation
Let us remind that in the BFKL case
 the structure function G can be represented as
\beq
G(x,Q^2)=\int (dM/2\pi)\exp(Mt+\alpha_s\chi(M)u)
\eeq
where $t=\log(Q^2/Q_0^2),u=\log(1/x)$.
In the BFKL case the coupling constant $\alpha_s$ is freezed. The integration contour is defined 
in the same way as in the DGLAP case.
For the Regge limit the integral can be taken using saddle point method.
The result is the energy independent saddle point $M_0=1/2$. It is easy to take
the Fourier transform and obtain
 according to the formulae above,
\beq
G(py,y^2)=\frac{1}{y^{5/2}}\exp(\alpha_s\chi(1/2)\log(py))
\label{b2}
\eeq
Here the BFKL anomalous dimension $\chi$ is given by
\beq
\chi(M) =\frac{3\alpha_s}{\pi}(2\psi(1)-\psi(M)-\psi(1-M))
\label{ad}
\eeq
with 
\beq
\chi(1/2)=\frac{12}{\pi}\log(2)
\label{as}
\eeq
 We obtain that the correlators are quickly
increasing with the distance and the oscillations are absent.
\par Within the resummation models of the ABF \cite{ABF} and Ciafaloni et
al \cite{Ciafoloni} it was argued that the behavior of the
structure functions in these models for the wide range of $x$ is
the same as in leading+plus nonleading order DGLAP. Numerically
this behavior is qualitatively quite similar to the leading order
DGLAP, except that the splitting functions is 15-30 $\%$ lower
than the leading order DGLAP, in the wide range of Bjorken $x$. At
quite low $x$  (say, $x\sim 10^{-6}$ for $Q^2\sim 4.5 $ GeV), the
DGLAP behavior of the kernel switches to BFKL type rise.(Note
\cite{Salam} that the switch point depends on $Q^2$ very weakly,
$x_c\sim 1/\sqrt{\alpha_s(Q^2)}$). This rise, although its
analytic form is unknown leads to the rapid rise of the correlator
with distance without oscillations \beq G \sim (py)^{\alpha_P-1}
\eeq Here $\alpha_P$ is the renormalized  perturbative Pomeron
slope of the resummation models, which is quite close (effectively
$\alpha_P\sim 1.25$). Note that the oscillations are absent
starting from sufficiently large py already in the DGLAP curve.

\section{Coherence length in QCD}

\par Let us now evaluate coherence length $l_c$  for the
total cross section of DIS evaluated within DGLAP aproximation.
We use here the same definition as B.Joffe length within parton
model \cite{Ioffe,IKL} and perform the inverse Fourier transform:
\beq
xG(x,Q^2)=2\nu\int^\infty_0dy^2\int^\infty_0duF(y^2,u)
\cos(\nu y^2/(2u)-xu), 
\label{inv} 
\eeq 
(the latter equation defining function F \cite{IKL}).
Let us carry on firstly  the integration over $y^2$. The relevant scale
for the integration over $y^2$ is $1/Q^2$. Thus within the LO DGLAP
approximation:   one may substitute $y^2\rightarrow 1/Q^2$,  so that 
the integral takes the form: 
\beq
xG(x,Q^2)=\int^\infty_0 4udu F(1/Q^2,u)\sin(1/2xu- xu) 
\label{inv1} 
\eeq
The coherence length is defined for given $Q^2,x$ as the typical;
$u$ in the integral. Such $u$ is either determined by the maximum of
the function F, or by the radius of oscillations in the sinus
multiplier. (For the another definition of $l_c$ and therefore different
dependence on x see Y.Kovchegov and M.Strikman \cite{KS}.)

\par $l_c$ within a parton model was determined by B.Ioffe who assumed
that  
\beq 
F(u)\rightarrow const 
\label{inv2} .
\eeq 
for $u\rightarrow \infty$.  If so  essential $u$  in the integral are completely 
determined by the oscillating multiplier , i.e. 
\beq 
u_I(x,Q^2)=1/x 
\label{inv3} 
\eeq 
The same behaviour is expected within DGLAP,  LO+NLO BFKL, resummation 
approximations. This is because in all approaches $l_c$ is effectively
life time of dipole with mass M where  $M^2=k_t^2/z(1-z)$. Here $k_t$ is 
transverse momentum of quark within dipole and $z$ is the fraction of dipole 
momentum carried by quark.  Thus  $l_c\approx q_o/Q^2+M^2$  and 
$Q^2\approx M^2$. Therefore $l_c\approx 1/2m_Nx$ .

\par  Near unitarity limit fast increase with energy of the coherence length should somewhat 
slow down. This is because the fast  increase of amplitude with energy leads to the related 
increase of probability of configurations of constituents with large $k_t$.  Let us consider for
definitness the case of longitudinally polarized photon where maximal important $k_t$
can be evaluated from pQCD formulae as:  $Q^2/k_t^2 1/x^{\mu(k_t^2)} \approx 1$  
So near unitarity limit:  $l_c \approx q_o/M^2 \propto 1/x^{1-\mu(k_{t}^2)}\ll 1/2m_{N}x$.  
 
\section{On the distorsion of dipole wave function near unitarity limit.}

\par Another consequence of the increase of the correlators with
the invariant length py is the deformation of the wave function of
the dipole near BDL. 

We found that fast increase with energy of LT contribution transforms in coordinate 
space into fast increase with distance of correlators of currents . The contribution of 
higher twist effects is increasing with x decrease even faster :$\propto 1/x^{n\mu(Q^2)}$ 
where n=1 for LT n=2 for first HT etc.  Performing the same Fourier transform of this formulae 
into coordinate space  as for LT term will produce even faster increase with distance of the 
contribution of HT effects as $\propto  (py) ^{1+n\mu(Q^2)}$. Thus dipole-target
interaction near unitarity limit is characterized by important role of long range interactions 
of dipole with target (which however are significantly smaller than B.Joffe length).  
This long-range target-dipole interaction effectively  leads to the interaction between the 
components of the dipole.  Indeed, at moderately small x  the contribution of dipole 
configurations with $z\approx 1/2$ is multiplied in the cross-section by the small factor -square 
of the dipole moment $d^2$.  Near BDL this suppression disappears and the
dominant configurations are $z\sim 1/2$.  Thus effective wave function of
the dipole changes significantly near the unitarity limit.

\section{Conclusion}
\par The main aim of the present paper was to demonstrate that the correlators of currents 
fastly increase with distance in pQCD .

\par We have also studied the issue of the space-time oscillations of the correlator of currents 
and found no oscillations in QCD contrary to the parton model where the correlator of currents 
is  $\sim \cos(py)$. Within DGLAP approximation oscillations are decreasing with distance 
because of decrease of coupling constant with virtuality.

\par We estimate coherent length near BDL as $l_c\propto x^{1-\lambda}$ .
\par We argue existence of the deformation of the dipole wave function near BDL 
resulting from the long range interaction between dipole and target \cite{BF}.
\par The most  suggesting consequence of our results is that in condensed matter
physics increase of correlations with distance usually leads to the phase transition.\cite{LL}
\par One of us (LF) is indebted to M.Strikman for the discussion of increase of correlators
with distance which led to the understanding of the impact of increase of gluon distribution 
on the coherence length.

\end{document}